\documentclass[12pt]{iopart}
\usepackage{epsf}
\begin{document}
\title{Testing Alternative Theories of Gravity using LISA}

\author{Clifford M Will\dag  
\footnote[3]{Permanent address:
 McDonnell Center for the Space Sciences, Department of
Physics, Washington University, St. Louis MO 63130, USA}
and Nicol\'as Yunes\ddag}
\address{\dag\ Groupe Gravitation Relativiste et Cosmologie
(GR$\varepsilon$CO)
Institut d'Astrophysique, 98 bis Boulevard Arago, 75014 Paris, France}
\address{\ddag\ Center for Gravitational Physics
and Geometry, Center for Gravitational Wave Physics,
Department of Physics,
The Pennsylvania State
University, University Park, PA 16802, USA}
\eads{\mailto{cmw@wuphys.wustl.edu}, \mailto{yunes@gravity.psu.edu}}

\date{today}

\begin{abstract}
We investigate the possible bounds which could be placed on
alternative theories of gravity using gravitational wave detection from
inspiralling compact binaries with the proposed LISA space interferometer.
Specifically, we estimate lower bounds on
the coupling parameter $\omega$ of scalar-tensor theories of
the Brans-Dicke type and
on the Compton wavelength of the graviton $\lambda_g$ in hypothetical massive
graviton theories.
In these theories, modifications of the gravitational 
radiation damping formulae or of the propagation of the waves translate
into a change in the phase evolution of the observed gravitational waveform.
We obtain the  bounds through the technique of matched filtering, employing
the LISA Sensitivity Curve Generator (SCG), available online.
For a neutron star inspiralling into a $10^3 \, M_\odot$ black hole in the
Virgo Cluster, in a two-year integration, we find a lower bound
$\omega > 3 \times 10^5$.  For lower-mass black holes, the bound could
be as large as $2 \times 10^6$.  The bound is independent of LISA arm
length, but is inversely proportional to the LISA position noise error.  Lower
bounds on the graviton Compton wavelength ranging from $10^{15}$ km to $5
\times 10^{16}$ km can be obtained from one-year
observations of massive binary black
hole inspirals at cosmological distances (3 Gpc), for masses ranging from
$10^4$ to $10^7 \, M_\odot$.  For the highest-mass systems ($10^7 \,
M_\odot$), the bound is
proportional to (LISA arm length)$^{1/2}$ and to (LISA acceleration
noise)$^{-1/2}$.  For the others, the bound is independent of these
parameters because of the dominance of white-dwarf confusion noise in the
relevant part of the frequency spectrum.  These bounds improve and extend
earlier work which used analytic formulae for the noise curves. 

\end{abstract}

\submitto{\CQG}

\maketitle



\section{Introduction and summary}

The Laser Interferometer Space Antenna (LISA) is a
gravitational-wave detector being designed for launch in the 2010-2015 time
frame \cite{danzmann}.  Consisting of a triangular array of three 
satellites orbiting the Sun
on an Earth-like orbit, it will use laser interferometry to open up the
low-frequency gravitational-wave window, to complement the high-frequency
window currently being explored by ground-based interferometers.  It is
expected to be able to observe waves from known binary star systems, from a
background of white-dwarf binaries, from inspirals of black holes and other
compact bodies into massive black holes, and possibly from phase transitions
in the early universe.  

LISA may also provide new and interesting tests of fundamental physics.  In
previous papers \cite{willbd,willmg,scharrewill}, we showed how observations
of waves from inspiralling compact binaries could place bounds on
alternative theories of gravity, such as theories of the scalar-tensor type
(e.g. Brans-Dicke theory), or theories with a massive graviton.  

In scalar-tensor theories,  
the phenomenon of dipole gravitational radiation modifies the damping of the
binary orbit, and thereby alters 
the evolution of the phasing of the received wave,
compared to what general relativity would predict.  We showed, for example,
that, for the
inspiral of a neutron star into a black hole of mass
$10^3\, {\rm M}_\odot$ at a distance of 50  Mpc, the lower bound on the
coupling parameter $\omega$ would be $2.4 \times 10^5$, with one year of
integration prior to the innermost stable orbit \cite{scharrewill}.  
The bound falls off with
increasing mass.

In massive graviton
theories, the wavelength-dependent propagation speed of the waves and the
resulting arrival-time offsets also modify the evolution of 
phasing of the received wave, compared to general relativity.  
For the inspiral of two $10^6 \, M_\odot$ black holes at 3 Gpc, we showed that
the lower bound on the graviton Compton wavelength $\lambda_g$ would be $5.4
\times 10^{16}$ km, over four orders of magnitude larger than the bound
inferred from solar-system dynamics.

Several developments have motivated us to revisit 
these bounds.  One is the dramatic improvement in the solar-system bound on
scalar-tensor gravity via tracking of the {\em Cassini} spacecraft 
\cite{cassini}.
The new bound is $\omega > 4 \times 10^4$.  Another is the increasing
interest in massive gravity theories, from the viewpoint both of
conventional modifications of general relativity \cite{babak1,babak2}, and
of multidimensional or brane-world theories \cite{branes}.   

The third
factor is the availability of an improved noise curve for the LISA
detector.  Earlier work employed an analytic approximation to LISA's
instrumental noise based on the LISA Phase A study \cite{LISAphaseA,cutler},
augmented by an estimate of white-dwarf confusion noise in the low-frequency
regime \cite{benderhils}.  The improved curves, based on work by Larson {\em et al.} and
Armstrong {\em et al.} \cite{larson,armstrong}, additionally 
take into account the
finite propagation time of the laser signals during the passage of the
gravitational wave, which is embodied in a transfer function that modifies
the response of LISA at high frequencies.  In addition, the curves are
available online in a ``Sensitivity Curve Generator'' (SCG) \cite{scgwww}, 
that permits the user to modify the 
parameters of LISA (arm length, position noise budget, etc) so as to explore
the capabilities of different hypothetical LISA instruments.
The main difference between the analytic curves used earlier and those from
the SCG, apart from the characteristic
oscillations at high frequency arising from the transfer function, is that
the amplitude noise level of the latter is roughly a factor of 3 higher 
(see figure 1).  
This will have an effect on some of the bounds we
report.

\begin{figure}[t]
\bigskip
\bigskip
\begin{center}
\epsfxsize=10cm
\leavevmode
\epsfbox{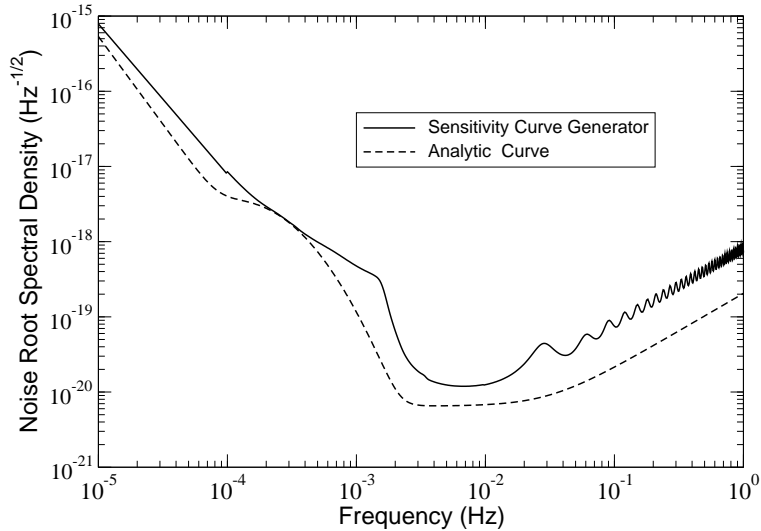}
\caption{\label{fig1} LISA root spectral noise density vs. frequency,
including white-dwarf confusion noise.  Shown is the nominal LISA curve from
the Sensitivity Curve Generator, and the analytic curve used in 
\cite{willmg,scharrewill}}
\end{center}
\end{figure}

We use the method of parameter estimation via matched filtering to estimate
bounds on alternative theories.  The idea of matched filtering is to 
cross-correlate
a theoretical template gravitational waveform against the output of the
detector.  The template may depend on a number of parameters, such as the
masses of the stars, and parameters associated with the theory of gravity.
One can then calculate, for a given set of parameters and for a given
spectrum of noise in the detector, how accurately the parameters inherent in
the true signal can be estimated 
\cite{3min,CutlerFlanagan,Finn,FinnChernoff,poissonwill}.

\begin{figure}
\bigskip
\bigskip
\begin{center}
\epsfxsize=10cm
\leavevmode
\epsfbox{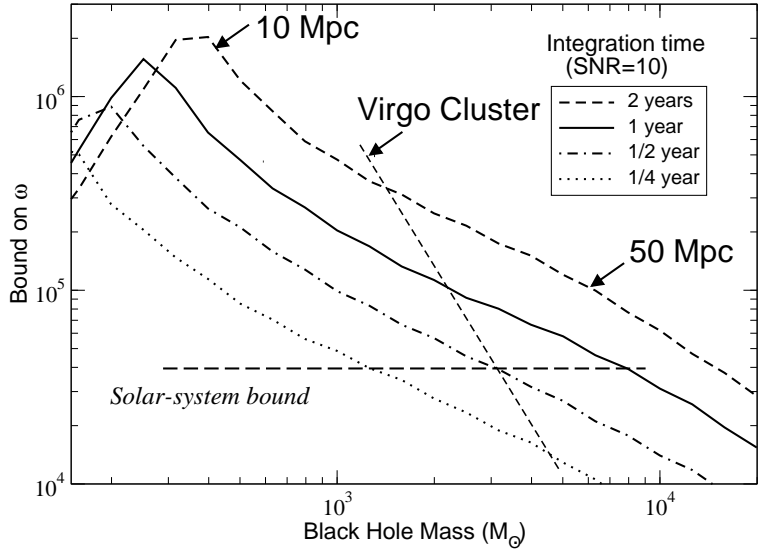}
\caption{\label{fig2} Bound on Brans-Dicke $\omega$ vs. black-hole mass,
assuming SNR$=10$}
\end{center}
\end{figure}

For testing scalar-tensor theories, the best system is binary inspiral of a
neutron star into a black hole.  We consider quasi-circular inspiral of
non-spinning bodies ending at the innermost stable circular orbit
(ISCO).  Assuming
detection with a signal-to-noise ratio (SNR) of 10, we find lower bounds on
$\omega$ shown in figure 2, for various integration times.  Also shown
are representative distances of the sources observed, along with the
solar-system bound from {\em Cassini}.  The sharp decrease in the bounds for
BH masses below a few hundred solar masses results from the fact that the
signal falls off the high-frequency end of the LISA sensitivity curve before
the system reaches the ISCO.  The bounds shown in figure 2 are only slightly
changed from our earlier bounds (figure 1 of \cite{scharrewill}).  For a fixed
SNR, the bounds depend on the shape of the noise curve, not on
its overall scale (the oscillations from the transfer function tend partially 
to
average out).  The change occurs in the distance at which a given source
can be detected, or in the bound obtained from a source at a given distance.  
Both have decreased by about a factor of three, consistent with the shift
observed in figure 1.  
We also find that, for a source at a given distance,
the bound is relatively insensitive to LISA arm length
but is inversely proportional to LISA position error.

For testing massive graviton theories, the best systems are massive black hole
binaries.  Again we consider quasi-circular inspiral of
non-spinning bodies ending at the innermost circular orbit.
For sources at a redshift $Z=1/2$, we find lower bounds on $\lambda_g$ shown
in figure 3.  The bounds are plotted for various total masses as a function
of the reduced mass parameter $\eta=m_1m_2/(m_1+m_2)^2$, which varies from
zero to $1/4$.  The curves terminate at the small mass ratio end where the
SNR drops below 10.   For sources at a given distance, the
bounds on $\lambda_g$ are weaker than those reported earlier \cite{willmg} by 
a factor of about $\sqrt{3}$ (square root because of the quadratic
dependence of the effect on $\lambda_g$), 
again reflecting the higher noise level generated by the SCG
compared to our earlier formula.  
For the highest mass systems, e.g. $10^7 \times 10^7 M_\odot$, the bound
is proportional to (LISA arm length)$^{1/2}$ and to (LISA acceleration
noise)$^{-1/2}$.  This dependence becomes progressively weaker with
decreasing mass, so that for $10^4 \times 10^4 M_\odot$
systems, the bound is independent of these parameters.  This is because 
the signals from high
mass systems reside at the low frequency end of the LISA noise curve, where
arm length or acceleration noise affect the noise in the expected fashion,
whereas the signals from
lower mass systems reside in the regime where the noise is dominated
by white-dwarf confusion noise.  

The bounds shown in figure 3 should be
compared to bounds inferred from the validity of {\it static} Newtonian
gravity over large distances, of $2.8 \times 10^{12}$ km from solar-system
dynamics \cite{talmadge}, and $6 \times 10^{19}$ km from cluster dynamics
\cite{particle}.

\begin{figure}
\bigskip
\bigskip
\begin{center}
\epsfxsize=10cm
\leavevmode
\epsfbox{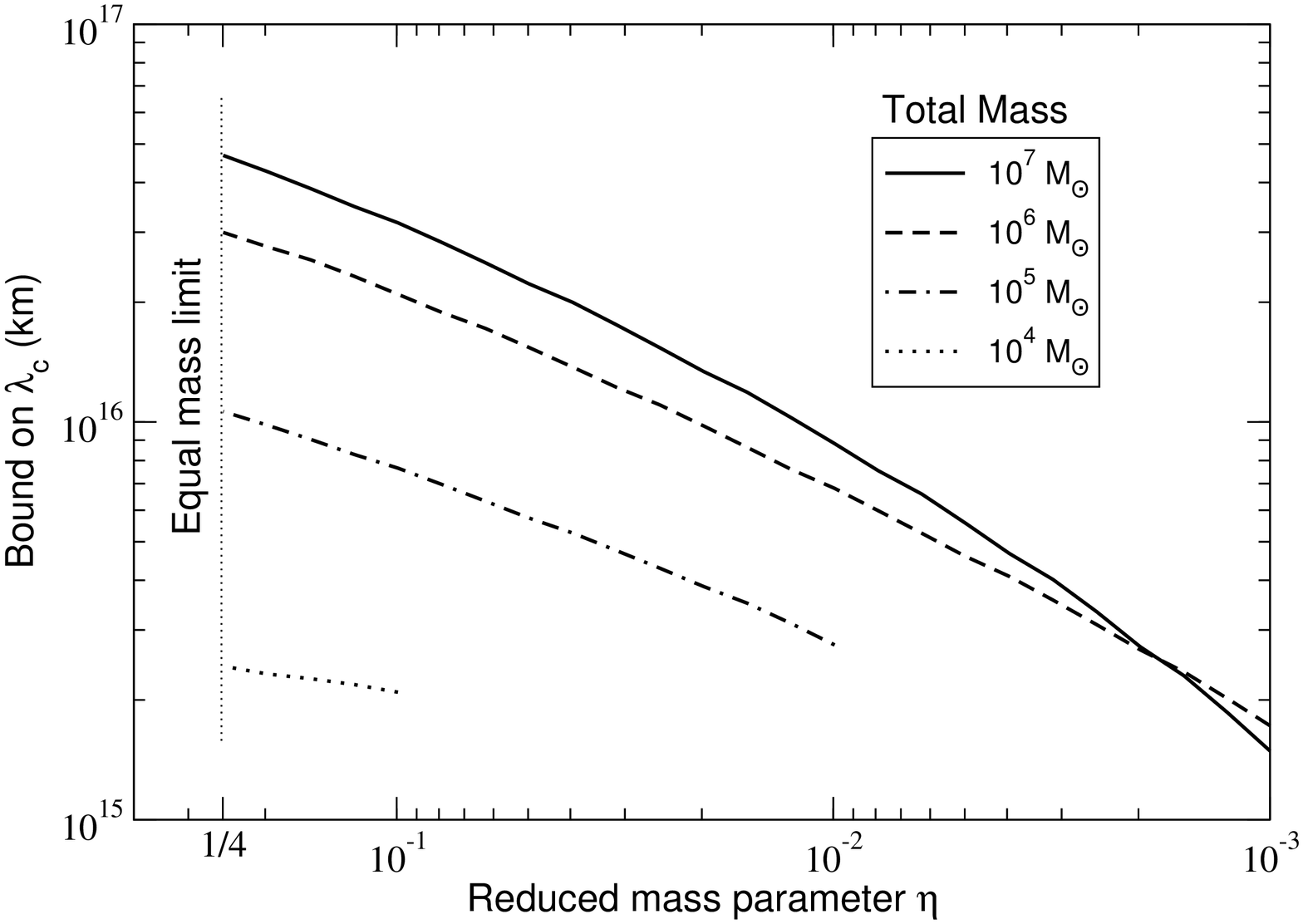}
\caption{\label{fig3} Bound on graviton Compton wavelength $\lambda_g$ vs.
$\eta=m_1m_2/(m_1+m_2)^2$, for massive black hole binaries at $Z=1/2$.}
\end{center}
\end{figure}

The remainder of this paper provides details.  In section 2 we review the
method of parameter estimation using matched filtering within general
relativity.  Section 3 treats scalar-tensor bounds and massive
graviton bounds.  Section 4 presents concluding remarks.

\section{Parameter estimation using matched filtering}

\subsection{Estimation in general relativity}

In this section we review briefly the technique used for estimating
parameters of inspiralling binaries using matched filtering, in the context
of general relativity.  Further details may be found in 
\cite{CutlerFlanagan,Finn,FinnChernoff,poissonwill}.
In matched filtering, a
template consisting of a theoretical gravitational waveform is
compared to the detector output.  The template that is a match with an
actual signal present with the noise will show a strong correlation with
the detector output and thereby will ``filter'' a signal out of 
background noise.  The template is represented
by $h(t)$, related to the spatial components of the radiative metric
perturbation far from the source.  We adopt 
the so-called ``restricted post-Newtonian
approximation'' for an inspiral orbit that is quasi-circular (that is, circular apart from the adiabatic 
decrease in separation), wherein we approximate $h(t) \approx
h_0(t)e^{i\Phi(t)}$, where $h_0(t)$ is a slowly varying wave amplitude
which depends on the 
wave polarization, the
location of the source on the sky, 
the detector orientation,  and the distance
to the source.  The wave phase, $\Phi(t)$, is a
function of the evolving orbital frequency.  
Because the process of matched-filtering is sensitive mostly to the phasing
of the wave, 
the amplitude $h_0(t)$ is 
assumed to be given by the lowest-order,
quadrupole approximation,  
while $\Phi(t)$ is taken to a suitably high post-Newtonian order.  The 
Fourier transform of $h(t)$ using the stationary phase
approximation is given by
\begin{equation}
\tilde{h}(f) = \left\{  \begin{array}{ll}
{\cal A} f^{-7/6} e^{i\Psi(f)} \,,  & 0<f<f_{\rm max} \,,\\
0  \,, & f>f_{\rm max} \,,
\end{array}  \right. 
\label{fourier}
\end{equation}
where $f_{\rm max}$ is the largest frequency for which the wave can be 
described by the restricted post-Newtonian approximation,  
often taken to correspond to waves 
emitted at the innermost stable orbit (ISCO) before the bodies 
plunge toward each other and merge.  A useful approximation for this
frequency is (we use units in which $G=c=1$)
\begin{equation}
f_{\rm max} = (6^{3/2} \pi m)^{-1},
\label{iscofreq}
\end{equation}
where $m$ is the total mass of the system.
After averaging over all angles, the amplitude $\cal A$ is given by
\begin{equation}
{\cal A} = \frac{1}{\sqrt{30} \pi^{2/3}} \frac {{\cal M}^{5/6}}{D_L} \,,
\label{amplitude}
\end{equation}
where  
$D_L$ is the luminosity
distance of the source, and ${\cal M} = \eta^{3/5}m$ is the ``chirp'' mass.
In general relativity, the phasing function $\Psi(f)$ is given, through 1.5
post-Newtonian (PN) order, for spinless bodies, by
\begin{eqnarray}
\Psi(f) &=& 2 \pi f t_c - \phi_c - \frac{\pi}{4}  \nonumber \\
&& + \frac{3}{128}u^{-5/3}
\left[1
+ \frac{20}{9}\left(
\frac{743}{336} + \frac{11}{4} \eta \right)\eta^{-2/5} u^{2/3}
- 16 \pi \eta^{-3/5} u \right] \,,
\label{psifinal}
\end{eqnarray}
where $u=\pi {\cal M} f$ ($u \sim v^3$) and $\phi_c$ is formally defined
as the phase of the wave at the time of coalescence, $t_c$.  (Terms through
3.5PN order are known \cite{lucwave} but will not be used here.)

By maximizing the correlation between a template waveform that depends on a set 
of parameters 
$\theta^a$  (for example, the chirp mass $\cal M$) and a measured signal, 
matched filtering provides 
a natural way to estimate the parameters of the signal and their errors. 
With a given detector noise spectral density 
$S_n(f)$ one defines the inner product 
between two signals $h_1(t)$ and $h_2(t)$ by
\begin{equation}
(h_1|h_2) \equiv 2 \int_0^{\infty} \frac{ {\tilde{h}_1}^*\tilde{h}_2 +
{\tilde{h}_2}^*\tilde{h}_1 }{S_n(f)}df \,,
\label{innerproduct}
\end{equation}
where $\tilde{h}_1(f)$ and $\tilde{h}_2(f)$ are the Fourier transforms 
of the respective gravitational waveforms
$h(t)$.  The signal-to-noise ratio (SNR) for a 
given $h$ is given by
\begin{equation}
\rho[h] \equiv  (h|h)^{1/2} \,.
\label{rho}
\end{equation}
One then defines the ``Fisher information matrix'' $\Gamma_{ab}$ 
with components given by
\begin{equation}
\Gamma_{ab} \equiv \left( \frac{\partial h}{\partial\theta^a} \mid
\frac{\partial
h}{\partial\theta^b} \right) \,,
\label{fisher}
\end{equation}
An estimate of the rms error, $\Delta\theta^a$, in measuring  
the parameter $\theta^a$ can then be 
calculated, in the limit of large SNR, by
taking the square root of the diagonal elements of the inverse of the
Fisher matrix,
\begin{equation}
\Delta\theta^a = \sqrt{\Sigma^{aa}} \,, \qquad  \Sigma = \Gamma^{-1} \,.
\label{errors}
\end{equation}
The correlation coefficients between two parameters $\theta^a$ 
and  $\theta^b$  are given by
\begin{equation}
c_{ab} = \Sigma^{ab}/\sqrt{\Sigma^{aa}\Sigma^{bb}} \,.
\label{correlations}
\end{equation}

In general relativity, the parameters to be estimated using the above
template
would be 
$\phi_c$, $f_0 t_c$, $\ln {\cal M}$, and $\ln \eta$, 
where 
$f_0$ is a fiducial frequency characteristic 
of the detector noise spectrum.  
The method then follows that 
used, for example,
in \cite{poissonwill}: combining equations (\ref{fourier}) and
(\ref{psifinal}) and calculating the partial derivatives $\partial {\tilde h}
/\partial \theta^a$ for the four listed parameters, we construct the
Fisher information matrix using equations (\ref{innerproduct}) and
(\ref{fisher}).  We then invert the information matrix and
evaluate the errors in the four parameters, along with the correlation
coefficients, notably between $\cal M$ and $\eta$.  
For the alternative Brans-Dicke
or massive graviton theories to be considered here, we will add a suitable
term to the phasing $\Psi(f)$, dependent upon an additional
parameter $\theta$.  We will
augment the dimension of the Fisher matrix by one, and estimate five parameters,
along with the correlation
coefficients between $\cal M$, $\eta$ and $\theta$.  Since the nominal value
of $\theta$ will be assumed to be zero (corresponding to
$\omega=\infty$ or $\lambda_g=\infty$), the error on $\theta$
will translate into a lower bound on $\omega$ or $\lambda_g$.  

\subsection{Space-based interferometers}

We consider space-based interferometers of the proposed LISA type, with 
a sensitive bandwidth 
between $10^{-5}$ and $1$ Hz, and an expected 
noise curve which can be expressed in terms of
an overall amplitude $S_0$, and a function of the ratio $x=f/f_0$:
\begin{equation}
S_n(f) = S_0 g(x) \,.
\label{Snscaling}
\end{equation}
In earlier work \cite{willmg,scharrewill} 
we adopted an analytic noise curve that 
included the LISA instrumental noise and an estimate 
of ``confusion noise'' from a population 
of galactic white-dwarf binaries \cite{LISAphaseA,cutler,benderhils} given by
\begin{eqnarray}
S_0 &=& 4.2 \times 10^{-41} \, {\rm Hz}^{-1} \,, \nonumber \\
f_0 &=& 10^{-3} {\rm Hz} \,, \nonumber \\
g(x) &=& \sqrt{10} x^{-14/3} +1+x^2/1000 
 +313.5x^{-(6.398+3.518\log_{10}x)} \,.
\end{eqnarray}
In this paper, we will retain the scaling using $S_0$ and $f_0$, which is
useful for expressing analytic results, but for $g(x)$ we will use values
obtained from data files available on line using the LISA ``Sensitivity Curve
Generator'' (see section \ref{sec:scg}).  

In terms of this scaling, the 
SNR is given, from equations  
(\ref{fourier}), (\ref{innerproduct}) and (\ref{rho}), by
\begin{equation}
\rho^2 = 4 |{\cal A}|^2 f_0^{-4/3} I(7) / S_0, 
\label{rhonew}
\end{equation}
where we define the integrals $I(q)$ by
\begin{equation}
I(q) = \int_{x_{\rm min}}^{x_{\rm max}} \frac{x^{-q/3}}{g(x)} dx \,,
\label{Iq}
\end{equation}
where  $ x_{\rm min} = f_{\rm min}/f_0$ and 
$ x_{\rm max} = f_{\rm max}/f_0$, corresponding 
to the minimum and maximum frequencies over which the detector will 
integrate .  In some  calculations, the maximum 
value of $f_{\rm max}$  corresponds to 
radiation emitted at the ISCO of the 
system, while in others, 
we can consider the effect of terminating 
observations sooner than this final orbit.  
The frequency $ f_{\rm min} $ corresponds to the 
gravitational-wave frequency  observed a time $T$ earlier, 
where for LISA-type systems, we will assume 
$T \le $ two years.  Using the quadrupole approximation for radiation
damping, 
which is an adequate approximation for this purpose,
one can relate the frequencies of gravitational radiation at
the beginning and end of any time interval $T$ by the expression
\begin{equation}
u_i = u_f \left ( 1 + {256 \over 5} {T \over {\cal M}} u_f^{8/3} 
	\right )^{-3/8} \,.
\label{timeinterval}
\end{equation}

The luminosity distance $D_L$ to which such sources can be
detected can be obtained from
equations (\ref{amplitude}) and (\ref{rhonew}), and related to 
source 
masses, detector characteristics and the SNR: 
\begin{eqnarray}
\fl
D_L = \sqrt{\frac{2}{15}} \frac{{\cal M}^{5/6}}{\rho} (\pi f_0)^{-2/3} \left( 
\frac{I(7)}{S_0} \right)^{1/2} \nonumber \\  
\fl
\qquad = 2.45 \, {\rm Gpc}  \left( \frac{m_{NS}}{1.4M_\odot} \right)^{1/2}
\left( \frac{m_{BH} }{10^4 M_\odot} \right)^{1/3}
\left( \frac{10}{\rho} \right)
\left( \frac{4.2 \times 10^{-41} }{S_0} \right)^{1/2}
\left( \frac{10^{-3} }{f_0} \right)^{2/3}
I(7)^{1/2} 
\nonumber \\
\fl
\qquad = 4.81 \, {\rm Gpc} (4 \eta)^{1/2} \left( \frac{m_{Tot}}{10^6 M_\odot} \right)^{5/6}
\left( \frac{10^4}{\rho} \right)
\left( \frac{4.2 \times 10^{-41} }{S_0} \right)^{1/2}
\left( \frac{10^{-3} }{f_0} \right)^{2/3}
I(7)^{1/2}
\,.
\label{lumdistance}
\end{eqnarray}

\subsection{Sensitivity curves for LISA}
\label{sec:scg}

\begin{figure}
\bigskip
\bigskip
\begin{center}
\epsfxsize=10cm
\leavevmode
\epsfbox{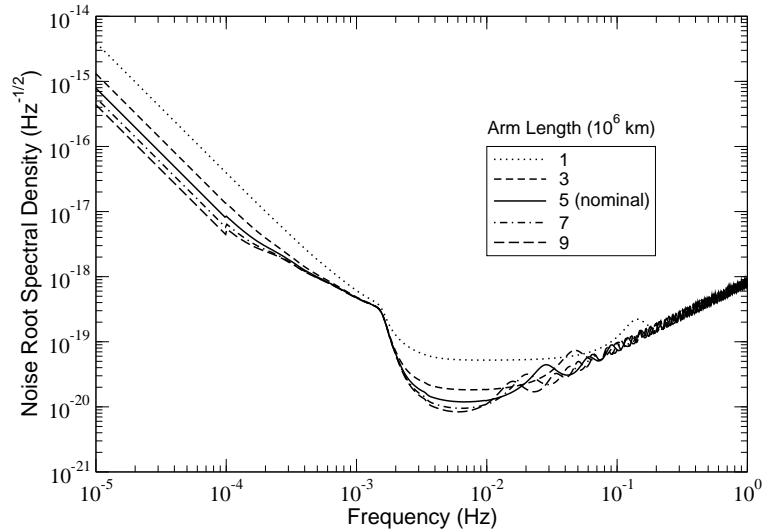}
\caption{\label{fig4} LISA noise root spectral density for varying arm length}
\end{center}

\end{figure}
\begin{figure}
\bigskip
\bigskip
\begin{center}
\epsfxsize=10cm
\leavevmode
\epsfbox{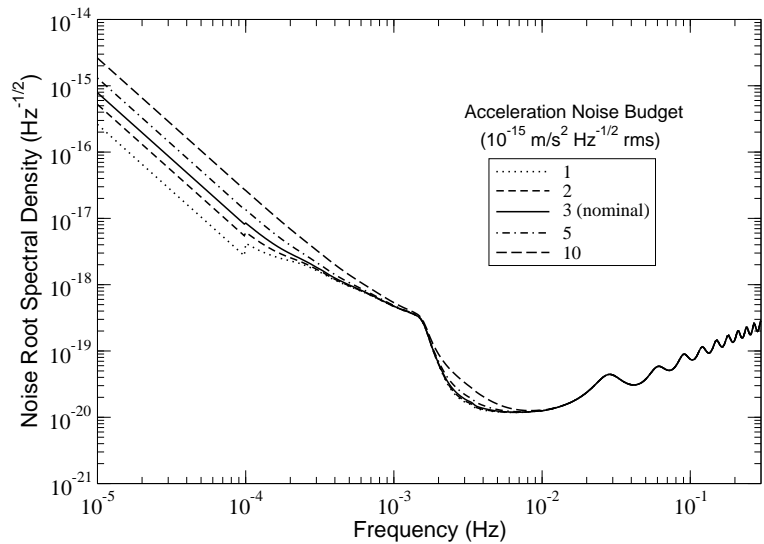}
\caption{\label{fig5} LISA noise root spectral density for varying
acceleration noise}
\end{center}
\end{figure}

\begin{figure}
\bigskip
\bigskip
\begin{center}
\epsfxsize=10cm
\leavevmode
\epsfbox{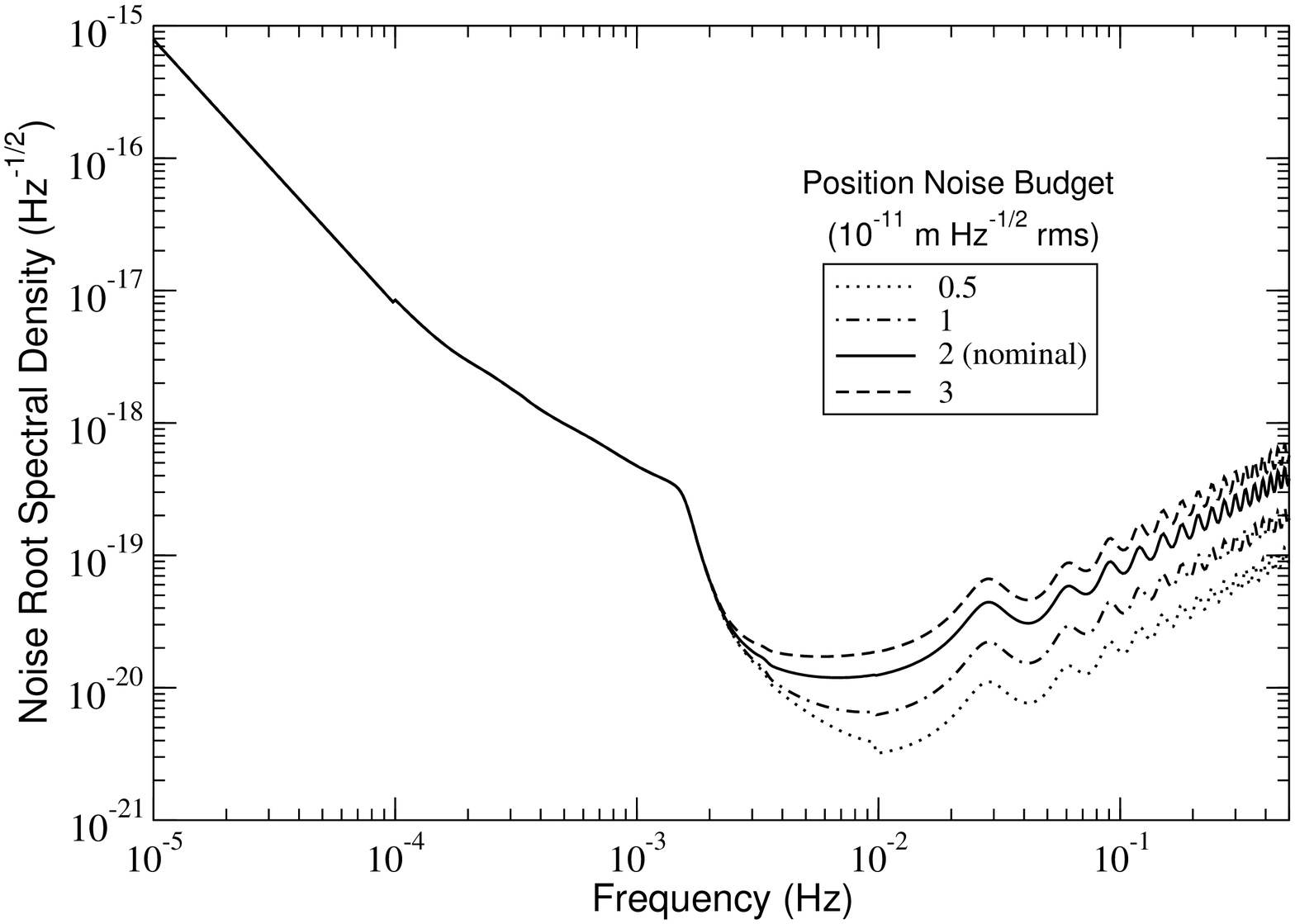}
\caption{\label{fig6}  LISA noise root spectral density for varying
position noise}
\end{center}
\end{figure}
In this paper, we shall adopt sensitivity curves for LISA developed
independently by Larson {\em et al.} \cite{larson} and Armstrong {\em et al.}
\cite{armstrong}.  The
two methods are in substantial agreement, and the former has been summarized
in the Sensitivity Curve Generator (SCG), available online
\cite{scgwww}.  The sensitivity curves incorporate sources of instrumental noise
in LISA, such as laser shot noise, acceleration noise, thermal noise, etc.,
coupled with a LISA transfer function which takes into account the effect of
the finite time of propagation of the laser beams during the passage of the
gravitational waves.  We assume the case where all three arms of LISA are of
equal length, and we assume
that averages over angles and polarizations have been done.
The SCG permits various choices to be made for LISA instrumental 
parameters, 
and has  an option to include 
an estimate for confusion noise resulting from a background
of galactic white-dwarf binaries
\cite{benderhils}; this background is included in our analyses.

For the baseline LISA proposal, the noise root spectral density ($h$ per
$\sqrt{Hz}$) is shown in figure 1, along with the analytic curve used in
earlier work.  
The oscillations at the high-frequency end result from the
transfer function.  
Notice that the SCG amplitude noise level is about a factor of 3 larger than
the analytic curve.  
This has the effect of weakening our bounds on $\omega$ and $\lambda_g$
relative to
earlier work.

In order to study the effect of varying LISA parameters on our tests of
alternative theories, we vary the arm length, the acceleration noise budget
and the position noise budget.  The resulting curves (including the WD
confusion noise) are shown in figures 4 -- 6.  
Varying arm length and acceleration
noise affects primarily the low-frequency noise, while varying the position
noise budget affects primarily the
high-frequency end.  Varying other parameters in the SCG, such as laser
power or wavelength, or telescope  diameter, has no sizable effect on the
noise curve or on the bounds.

\section{Testing alternative theories of gravity}

\subsection{Bounding scalar-tensor theories of the Brans-Dicke type}

One of the most striking differences between scalar-tensor theories 
and general relativity is the prediction of {\it dipole} gravitational
radiation.  The source of such radiation is the difference between the
self-gravitational binding energy per unit mass between the two bodies in a
binary system, as encoded in a ``sensitivity'' $s$ (technically, the sensitivity
of the body's total mass to variations in the locally measured value of the
gravitational constant).  For black holes, $s \equiv 0.5$, while for neutron
stars, $s \sim 0.1 - 0.2$, depending on the equation of state.  In
principle, dipole radiation can be a strong effect, because it depends on
two powers of velocity {\it fewer} than quadrupole radiation.  This
additional source of energy flux will alter the inspiral orbit of the
binary, and thus will modify the evolution of the waveform phasing.  

To sufficient accuracy, this can be taken into account in our phasing
function $\Psi(f)$ by adding the following term 
to the general relativistic formula, equation (\ref{psifinal}):  
\begin{equation}
\delta \Psi(f) = \frac{3}{128}u^{-5/3} \left ( -\frac{5}{84} \frac{{\cal
S}^2}{\omega} \eta^{2/5} u^{-2/3} \right ) \,,
\label{bdterm}
\end{equation}
where ${\cal S} = s_1-s_2$ (see \cite{willbd,scharrewill} for details).  
In doing so, we are assuming that $\omega \gg 1$, to be consistent
with the solar-system bound $\omega > 40,000$, 
so that the $O(1/\omega)$ corrections
to the general relativistic terms in equation (\ref{psifinal}) 
may be ignored compared to
the dipole term of equation (\ref{bdterm}).  
Notice that, compared to the leading term ``1'' in equation
(\ref{psifinal}), the dipole term is $O(u^{-2/3}) \sim O(1/v^2)$.

Because $s=0.5$ for black holes, binary black holes do not emit dipole
radiation at all.  For neutron stars, $s$ is a relatively weak function of
mass, and so for binary neutron stars with masses near $1.4 \, M_\odot$,
dipole radiation is suppressed by symmetry.  The only promising sources,
then, are  mixed, such as black-hole neutron-star systems 
(black-hole white-dwarf
systems were discussed in \cite{scharrewill}).   

\begin{table}[t]
\begin{center}
\leavevmode
\begin{tabular}{rrrrrrrrr}
\hline
$m_{BH} $&$\Delta t_c$&$\Delta \Phi_c$&$\Delta {\cal M}/{\cal M}$
&$\Delta\eta / \eta$&Bound  
& $c_{\eta {\cal M}}$ & $c_{{\cal M} \tilde{b}}$ & $c_{\eta\tilde{b}}$ 
\\
$(M_\odot)$&\hfil(s)\hfil&&\hfil (\%)\hfil &\hfil (\%)\hfil &\hfil
on $\omega$\hfil &&& 
\\ \hline

1000&	3.94&	18.3&	.000222&.1034&	203772&	.886&	-.994&	-.929
\\

5000&	3.78&	11.5&	.000528&.0246&	58020&	.970&	-.998& -.954
\\

10000&	5.27&	12.6&	.000739&.0174&	31062&	.976&	-.997&	-.957
\\
\hline

\end{tabular}

\caption{Estimated parameter errors for $1.4 \, M_\odot$ NS-MBH 
systems in Brans-Dicke theory:  SNR = 10, integration 
time
is one year prior to the ISCO, neutron star sensitivity $s_{NS} = 0.2$. 
} 
\label{tab:tablebd} 
\end{center}
\end{table}

We now carry out the parameter estimation calculation for the parameters
$\phi_c$, $f_0t_c$, $\ln {\cal M}$, $\ln \eta$, and $b \equiv 5{\cal
S}^2\eta^{2/5}/48\omega$ ($b \to 0$ in the GR limit).  We consider a
neutron star of mass $1.4 \, M_\odot$ in a quasicircular inspiral orbit 
around a black hole of mass $m_{BH}$.  We adopt the value $s_{NS} = 0.2$.  The
results of the parameter estimation for various black-hole masses are shown
in table 1, and the bounds on $\omega$ as a function of black hole mass
and for various integration times 
are shown in figure 2.    For black holes less massive than a few thousand
solar masses, the bounds could exceed the current solar system bound.
figure 7 shows, for a $10^3 \, M_\odot$ black hole, the bound that could be
achieved for sources at various redshifts, and as a function of the LISA
position error budget ($2 \times 10^{-11} \, {\rm m}/\sqrt{\rm Hz}$ is the
nominal value).  
Since $\Delta b \propto 1/\rho$, the bound on $\omega$ should be inversely
proportional
to the position error budget, as confirmed in figure 7.
The bounds can also be shown to be relatively insensitive
to LISA arm length or acceleration noise.  
These results are consistent with figures 4 -- 6, since the
relevant sources are concentrated at the high-frequency end of the LISA
noise spectrum, where the noise spectral density is affected by position
errors but not by arm length or acceleration noise.  
These bounds for sources at a set distance,
are roughly a factor 3 weaker
than the bounds estimated in \cite{scharrewill}, as expected from the
overall shift shown in figure 1.

\begin{figure}
\bigskip
\bigskip
\begin{center}
\epsfxsize=10cm
\leavevmode
\epsfbox{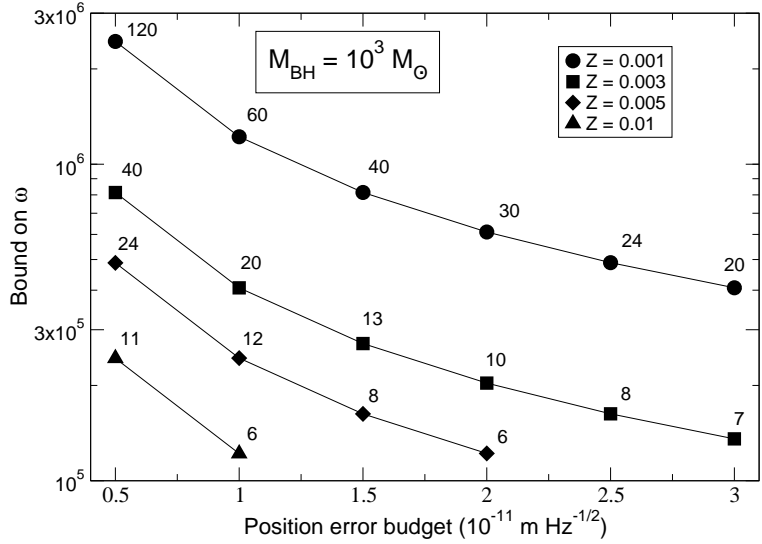}
\caption{\label{fig7} Bound on $\omega$ vs. LISA position error, for
sources at various redshifts.  Shown are
values of the SNR for representative points.}
\end{center}
\end{figure}

\subsection{Bounding massive graviton theories}

Notwithstanding statements in the literature forbidding theories of gravity
with a massive graviton \cite{zakharov,vandam}, 
such theories have attracted considerable
recent interest, from the point of view of conventional modifications of
classical GR \cite{visser,babak1,babak2}, and from considerations of extra
dimensions \cite{branes}.  
To the extent that a candidate massive graviton theory
merges smoothly with GR in the limit that the graviton mass $m_g$ vanishes or
its Compton wavelength $\lambda_g \to \infty$, one might reasonably expect
corrections to GR results for the intrinsic behaviour of a system
to be of order $(R/\lambda_g)^2$, where $R$ is the
characteristic size of the system.  Since solar system measurements already
constrain $\lambda_g > 10^{12}$ km, such corrections can be ignored for
inspiralling compact binaries.   The dominant effect of a massive graviton
is via the propagation of the waves: lower frequency waves propagate more
slowly than higher frequency waves.  For binary inspiral
seen at cosmological distances, this
wavelength dependent speed difference can result in a significant 
cumulative distortion
of the apparent phasing of the observed chirp gravitational-wave signal.  The
effect is to add the following term
to the general relativistic formula, equation (\ref{psifinal}):
\begin{equation}
\delta \Psi(f) = \frac{3}{128}u^{-5/3} \left ( -\frac{128}{3} 
\frac{{\pi^2 D \cal M}}{\lambda_g^2 (1+Z)} u^{2/3} \right ) \,,
\label{gravitonterm}
\end{equation}
where $Z$ is the redshift of the source, and 
$D=(1+Z)\int_{t_e}^{t_0} [a(t)/a(t_0)]dt$, where $a(t)$ is the cosmological
scale factor, and $t_e$ and $t_0$ are the times of emission and reception of
the gravitational-wave (see \cite{willmg} for details).  
The ``distance'' $D$ arises from the
wavelength-dependent propagation of the gravitational wave signal.  It is
related to the normal luminosity distance $D_L$ by $D/D_L =
[1+(2+Z)(1+Z+\sqrt{1+Z})]/5(1+Z)^2$ in a spatially flat, matter dominated
universe.

\begin{table}[t]
\begin{center}
\leavevmode
\begin{tabular}{rrrrrrrrrrr}
\hline
$m_{1} $&$m_{2} $&SNR&$\Delta t_c$&$\Delta \Phi_c$&$\Delta {\cal M}/{\cal M}$
&$\Delta\eta / \eta$&Bound on  
& $c_{\eta {\cal M}}$ & $c_{{\cal M} \beta}$ & $c_{\eta\beta}$ 
\\ 
$(M_\odot)$&$(M_\odot)$&&(s)&&\hfil (\%)\hfil &\hfil (\%)\hfil &\hfil
$\lambda_g$ (km)&\hfil  
\\ \hline

$10^7$&	$10^7$&	1023&	31.3&	.119&	.0316&	.945&	$4.8 \times 10^{16}$&
-.979&	-.991&	.997
\\

$10^6$& $10^6$&	984&	1.65&	.048&	.0050&	.276&	$3.1 \times 10^{16}$&
-.965&	-.988&	.993 
\\

$10^5$& $10^5$&	871&	.202&	.034&	.0015&	.140&	$1.5 \times 10^{16}$&
-.953&	-.986&	.987 
\\

$10^4$& $10^4$&	128&	.756&	.286&	.0013&	.441&	$0.4 \times 10^{16}$&
-.957&	-.988&	.989
\\
\hline
\end{tabular}

\caption{Estimated parameter errors for binary MBH 
systems at 3 Gpc in massive graviton theories:  integration 
time
is one year prior to the ISCO.  
} 
\label{tab:tablemg} 
\end{center}
\end{table}

We now carry out the parameter estimation calculation for the parameters
$\phi_c$, $f_0t_c$, $\ln {\cal M}$, $\ln \eta$, and 
$\beta=\pi^2 D{\cal M}/\lambda_g^2 (1+Z)$ 
($\beta \to 0$ in the GR limit).  We consider massive
binary black hole systems without spin, 
in quasicircular inspiral orbits, for year-long
integration times leading to the
ISCO.  The sources are assumed to be at 3 Gpc, with a Hubble constant of 70
km/s/Mpc, and a spatially flat universe.  The
results of the parameter estimation for various black-hole masses are shown
in table 2.  The bounds on $\lambda_g$ as a function of black hole mass
for sources at $Z=1/2$ are shown in figure 3.

It is useful to note that, in the large $\rho$ limit, all errors such as
$\Delta \beta$
are inversely proportional to the SNR $\rho$. 
Defining $B^{1/2}\equiv \rho (\Sigma^{\beta\beta})^{1/2}$, 
viewing $\Delta \beta$ as an upper bound on $\beta$, and combining
this definition with equations (\ref{amplitude}), (\ref{errors})
and (\ref{rhonew}) we obtain 
an expression for the
{\it lower} bound on $\lambda_g$:
\begin{equation}
\lambda_g >  \left ( {2 \over 15}{I(7) \over S_0} \right )^{1/4} \left ( {D
\over (1+Z)D_L} \right )^{1/2} {{\pi^{2/3} {\cal M}^{11/12}} \over
{f_0^{1/3} B^{1/4}}} \,.
\label{lambdabound}
\end{equation}
Note that the bound on $\lambda_g$ depends only weakly on distance, via
the $Z$ dependence of the factor $[D/(1+Z)D_L]^{1/2}$, which varies
from unity at $Z=0$ to 0.68 at $Z=1/2$.  This is because, while
the signal strength and hence the accuracy fall with distance, the
size of the arrival-time effect increases with distance.  Otherwise,
the bound depends only on the chirp mass and on detector
noise characteristics.

\begin{figure}
\bigskip
\bigskip
\begin{center}
\epsfxsize=10cm
\leavevmode
\epsfbox{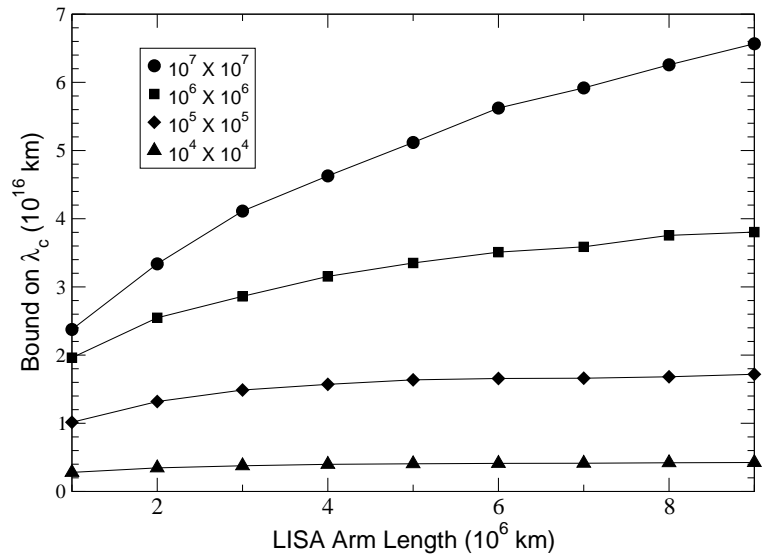}
\caption{\label{fig8} Bound on graviton compton wavelength $\lambda_g$ vs.
LISA arm length for sources at $Z=1/2$; $5 \times 10^6$ km is the baseline. }
\end{center}
\end{figure}

\begin{figure}
\bigskip
\bigskip
\begin{center}
\epsfxsize=10cm
\leavevmode
\epsfbox{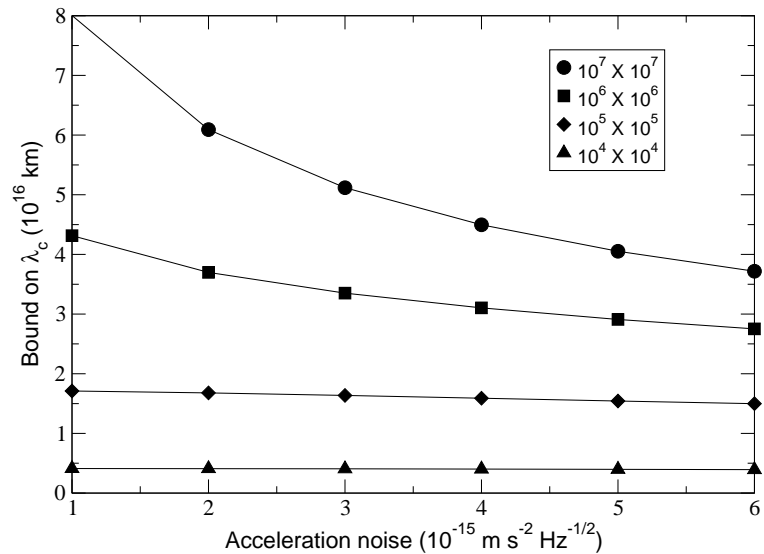}
\caption{\label{fig9} Bound on graviton compton wavelength $\lambda_g$ vs.
LISA acceleration noise for sources at $Z=1/2$; $3 \times 10^{-15} \, {\rm
m} \, {\rm s}^{-2} \, {\rm Hz}^{-1/2}$ is the baseline.}
\end{center}
\end{figure}

The resulting bounds are in the range between $10^{15}$ and several $\times
10^{16}$ km.  Figures 8 and 9 show the dependence of the bounds 
on LISA arm length $L$ and acceleration noise $A$.
For the highest mass systems, the bounds vary as $L^{1/2}$, and
$A^{-1/2}$
because
the noise varies as $L^{-1}$ and $A$, 
while for the lower mass systems, there is little
effect because the relevant signal is in the regime where WD confusion
noise dominates over instrumental noise.  

\section{Discussion}

We have used up-to-date LISA sensitivity curves from the SCG to refine the
bounds that could be placed on alternative theories of gravity.

For scalar-tensor theory, the bound on $\omega$ could exceed that from the
best current solar-system measurement, using the {\em Cassini} spacecraft, by
factors of 10 and higher.  For neutron star inspiral into a $400 \, M_\odot$
black hole at 10 Mpc and a two-year integration, the bound could reach 2
million.  

On the other hand, a number of alternative bounds on scalar-tensor gravity
may reach levels competitive with these bounds on a comparable timescale.
The gyroscope experiment Gravity Probe-B, set for launch in April 2004,
anticipates placing a bound $\omega > 10^5$ via measurement of the geodetic
precession of gyroscopes orbiting in the curved spacetime around
the Earth.  Future space optical interferometery missions, such as GAIA,
could reach comparable levels by measuring the deflection of light to
microarcsecond precision.  GAIA is planned for launch around the same 
time (2010) as
LISA.  The best hope for a dramatic improvement in $\omega$
bounds comes from the recently analysed binary pulsar system PSR J1141-6545,
in which the companion is most likely a white dwarf \cite{bailes}.
Because of the asymmetry between the neutron star and white dwarf
sensitivities (0.2 vs. $10^{-4}$), dipole gravitational radiation is
significant, and a measurement of the rate of orbital decay ${\dot P}_b$ in
agreement with GR at the one percent level could bound $\omega$ by as high
as $10^6$ \cite{gilles04,wiaux}.  Such a result could possibly
be reached in a decade,
the same timescale as LISA.

A major uncertainty in our proposed bound using LISA
is the likelihood of observation of relatively nearby
inspirals of a neutron star into intermediate mass black holes.  Miller
\cite{miller02} has estimated the rate of inspirals of intermediate-mass
binary black hole systems in globular clusters.  For a $10 \, M_\odot$
black hole inspiraling into a $100 \, M_\odot$ black hole, Miller
estimates yields a rate of one every 250 years, for one-year
integrations.   However one of us \cite{willimbh} has used the SCG to
estimate a rate $7 \times 10^5$ times lower.
Whether these estimates apply to the neutron-star inspirals
needed for scalar-tensor bounds is an open question at present.
Although the rate may be low from the point of view of gravitational-wave
astronomy, it is useful to point out that, to test alternative
theories, a single serendipitous discovery is all it takes: witness
the Hulse-Taylor binary pulsar.

Other methods have been suggested for bounding the graviton mass.  We
earlier showed that observations of binary inspiral using ground-based
detectors of the advanced LIGO type could place a bound of several times
$10^{12}$ km, comparable to the solar-system bound \cite{willmg}.  
Sutton and Finn \cite{sutton1,sutton2} showed that, in a simple class of
linearized massive graviton models, a bound comparable to or slightly better
than the solar-system
bound could be placed on $\lambda_g$ using binary pulsar data.
Cutler, Hiscock and Larson \cite{chl03}
studied the effect of massive gravity on the observed
gravitational-wave phase of known binary stars, compared to the orbital phase
inferred from light measurements and showed that LISA observations could
place bounds on $\lambda_g$ from 5 to 50 times larger than the current
solar-system bound.

\section*{Acknowledgments}

We are grateful to Shane Larson for useful discussions.
This work is supported in part by the National Aeronautics and Space
Administration, grant no NAG 5-10186 at Washington University,
and by the National Science Foundation,
grant no PHY 00-96522 at Washington University, and grant no
PHY-0245649 and cooperative agreement no PHY-0114375 at Pennsylvania
State University.  CW is grateful for the hospitality of the Institut
d'Astrophysique de  Paris during the academic year 2003-04.  NY 
thanks the Washington University Gravity Group and Physics Department
for 
their support. 

\end{document}